\documentstyle[aps,epsf,pre]{revtex}
\begin{document}

\hsize\textwidth\columnwidth\hsize\csname @twocolumnfalse\endcsname
\title{Floppy modes and the free energy:\\
              Rigidity and connectivity percolation on Bethe Lattices
          }    
\author{P.M. Duxbury, D.J. Jacobs, M.F. Thorpe }
\address{Department of Physics and Astronomy\\
     and Center for Fundamental Materials Research\\
         Michigan State University,
         East Lansing, Michigan 48824-1116}

\author{and C. Moukarzel}
\address{Instituto de F{\'{i}}sica\\
           Univesidade Federal Fluminense\\
           24210-340 Niter{\'{o}}i, Rio de Janeiro, Brasil}

\maketitle
\begin{abstract}
We show that  negative of the number of {\it floppy modes} behaves 
as a {\it free energy} for both connectivity and rigidity percolation,
and we illustrate this result using Bethe lattices.
The rigidity transition on Bethe lattices is found to be 
first order at a bond concentration close
to that predicted by Maxwell constraint counting.  
We calculate the probability of a bond being on the infinite cluster 
and also on the overconstrained part of the infinite cluster, and show 
how 
a {\it specific heat }  can be defined
as the second derivative of the free energy. 
We demonstrate that the Bethe lattice 
solution is equivalent to that of the random bond model, where points 
are joined randomly (with equal probability at all length scales)
 to have a given coordination, and then subsequently bonds 
are randomly removed.  
\end{abstract}


\twocolumn

\section{INTRODUCTION}
\label{Sec:I}
 
Connectivity percolation on Bethe lattices or infinite statistically
homogeneous Cayley trees was thoroughly analyzed by Fisher and Essam,\cite{FE}
and provides a useful model for percolation.  Rigidity on Cayley trees was
first studied by Moukarzel, Duxbury and Leath;\cite{MDL} henceforth referred
to as MDL.  This model has $g$ degrees of freedom per site, with $g=1$
corresponding to connectivity percolation being a special case, and so may be
regarded as an extension of the work of Fisher and Essam.\cite{FE} In this
paper, we develop a free energy for the rigidity percolation problem, and as
an example of its use, show how to locate the bulk rigidity transition on the
Bethe lattice.  We use the term {\it Bethe lattice}, rather than {\it Cayley
tree}, to emphasize the bulk behavior of lattices containing no loops away
from the boundary.\cite{FE,T1}

\begin{center}
\epsfxsize=2.0in          
\leavevmode\epsfbox{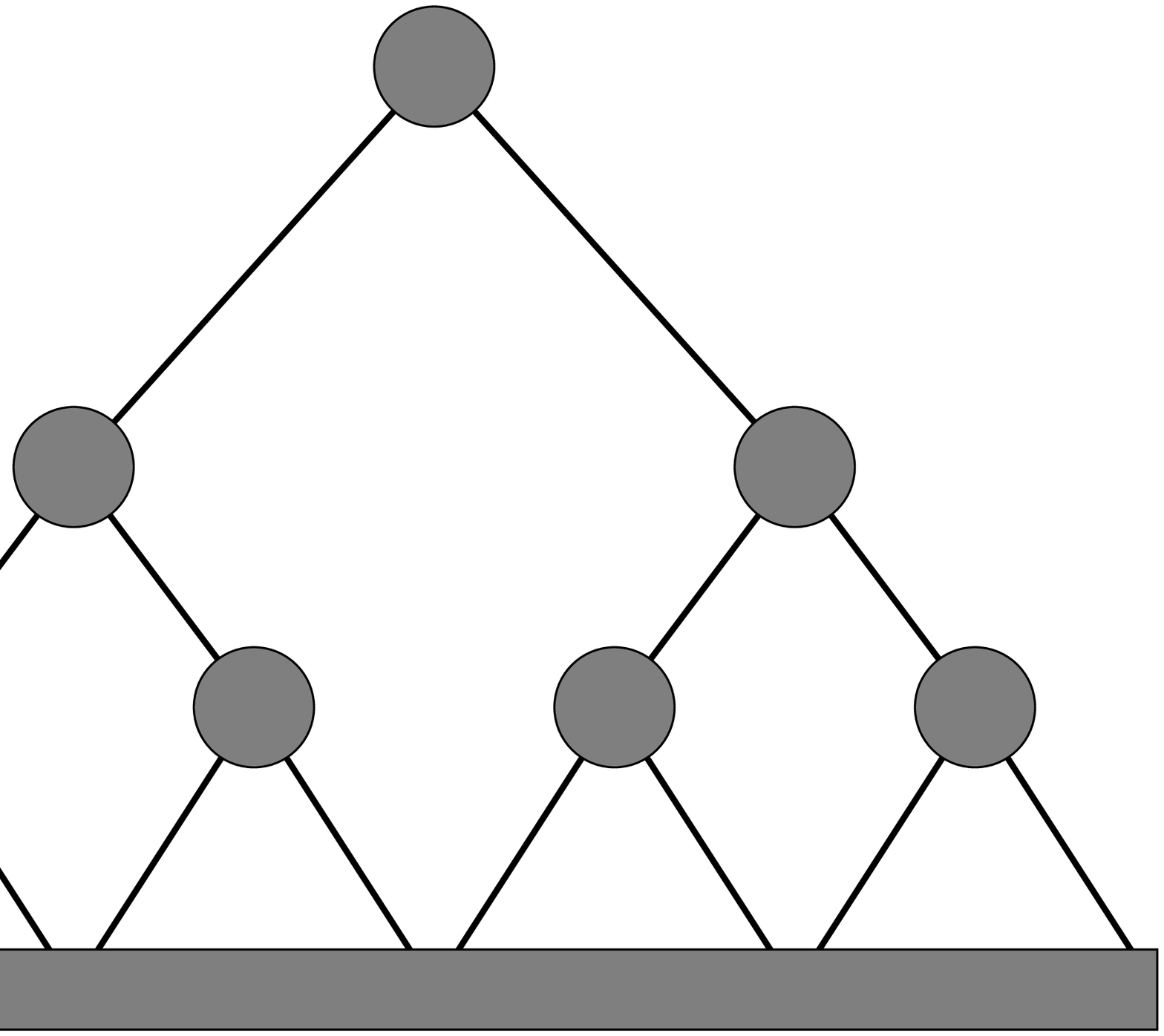}
\end{center}
\begin{figure}
\caption{
Showing a tree with a single bar $b=1$ along the $z=3$ bonds from each
 site.  The bonds are present with probability $p$.  Each site, shown as a
shaded circle, has $g$ degrees of freedom.  Only the first few levels from the
rigid busbar are shown.}
\label{graph1}
\end{figure}

We have previously suggested that a free energy can be defined as the negative
of the number of floppy modes.\cite{T2} In this paper we prove this assertion
for the case of random bond dilution in a general lattice with $g$ degrees of
freedom per site, and with $z$ nearest neighbors, connected by $b=1$ bars.
For $g=1$, this gives {\it connectivity} percolation.  For $g=d$, this
corresponds to the central force model \cite{SF,STG}, where $g=2$ in two
dimensions and $g=6$ in three dimensions.  For bodies joined by bars,
$g=d(d+1)/2$ in a $d$ dimensional space.\cite{MDL}

We prove that this free energy has the correct convexity property and hence a
{\it specific heat} can be defined from the second derivative.  This
definition of the free energy reduces to the known result for connectivity
percolation when $g=1$, where it becomes the negative of the total number of
isolated clusters.\cite{FK} This latter result can either be found
directly,\cite{FK} or as the $s \rightarrow 1$ limit of the $s$ state Potts
model.\cite{P} For rigidity, this free energy allows us to locate the bulk
transition on the Cayley tree, and we refer to this as the Bethe lattice
solution.\cite{T1} We confirm that for $b=1$ and $g \ge 2$ the rigidity
transition is always first order on the Bethe lattice, whereas of course it is
second order for connectivity percolation where $g = 1$.  On the triangular
lattice, rigidity percolation is second order.\cite{JT,JT2,MD}, whereas it is
first order on the Bethe lattice.  The nature of the rigidity transition is a
subtle question, depending on the network, and this has not always been taken
into account.\cite{O} For connectivity percolation, the transition is always
second order.\cite{FE,SA}

MDL developed a general solution, using a transfer matrix technique, for a
network of rigid bodies, each with $g$ degrees of freedom generically
connected by $b$ bars, as sketched in Fig.~1.
 Initially each rigid body has $z$ neighbors, and the rigidity of the network
is studied as the bonds are randomly removed.  For $b \le g$, this type of
network is always floppy, unless the tree is attached to a rigid surface
(busbar), in which case rigidity may or may not propagate away from the
busbar, depending on the degree of dilution.

We amplify the work of MDL by using the free energy and the associated Maxwell
construction,\cite{M} to remove boundary constraints in the analysis of bond
diluted Cayley trees, and hence obtain the bulk or Bethe lattice
solution. This constraint equation leads to a {\it higher} rigidity threshold
than that found in MDL.  In analogy with thermally driven thermodynamic
transitions, we interpret the threshold found in MDL as a {\it spinodal} point
and use $p_s$ when referring to it in this paper.  We use $p_c$ to refer to
the {\it bulk} threshold.

The layout of this paper is as follows.  In the next section (Sec.~II) we
develop results applicable to general bond-diluted lattices, namely: the proof
that the negative of the number of floppy modes acts as a free energy for
studying bond-diluted connectivity and rigidity percolation, and we derive the
associated Maxwell construction\cite{M} or consistency condition.
 We review the main equations from MDL for the Cayley tree, and evaluate the
probability $P_{\infty}$ of being in the infinite rigid cluster and also the
probability $P_{ov}$ of being in the overconstrained part of the infinite
cluster; both defined in the bulk far from the busbar.
 
 In Sec.~III, we discuss the {\it random bond model},\cite{JT3} where $z$
bonds join each site with other sites, regardless of distance, and then these
bonds are randomly diluted. This model is solved using the integer algorithm,
the pebble game,\cite{JT,JT2,JH} for the connectivity case $g=1$, $z=3$ and
the rigidity case $g=2$, $z=6$.  We demonstrate that in the thermodynamic
limit, the random bond model is equivalent to a Bethe lattice.
 
 In Sec.~IV, we show that for $g=1$, the familiar connectivity case is
recovered.  This analysis is examined for a general $z$, specializing to $z=3$
to illustrate the detailed behavior for connectivity percolation.
 
 In Sec.~V, we give detailed results for a Bethe lattice with $g=2$ and $z=6$
as an example of a first-order rigidity transition.  We apply the consistency
condition to locate the first order transition in the bulk, and show that the
results are equivalent to the random bond model. We also give results for the
probability $P_{\infty}$ of being on the infinite cluster and the probability
$P_{ov}$ of being on its overconstrained part.  We show that the results for
the number of floppy are equivalent to the $g=2$ random bond model.  We
present a table summarizing important results for a range of values of $g$ and
$z$.
 
 In Sec.~VI, we show how the limit $z \rightarrow \infty$ can be taken.  We
show when a further limit $g \rightarrow \infty$ is also taken, we recover the
{\it Maxwell constraint counting} result, which we therefore identified as the
mean field theory for rigidity.
  
  Throughout this paper we focus our attention on bond dilution where the
number of bars $b=1$.  Results for $b \ge 2$ and$/$or for site dilution can
probably be obtained in a similar manner.

\section{Formalism}
\label{Sec:II}
\subsection{General}
\label{Sec:IIa}

In this subsection we develop a
 free energy for both connectivity and rigidity problems on general lattices,
and show that this leads to a consistency constraint on the number of floppy
modes that is useful in locating first order transitions.  These results
 are applied to Cayley trees in the next sections.
 
  A constraint counting method (also called {\it{Maxwell counting}}) has been
very useful as a conceptual tool in understanding rigidity, and in particular
 the rigidity of glasses.\cite{T2,P2,HT,CT}
 The Maxwell counting approach centers on the number of floppy modes, where we
define $f(p)$ as the average number of floppy modes per degree of freedom.  We
consider generic networks where each site has $g$ degrees of freedom with
central forces that connect to $z$ neighbors. Bonds are randomly present with
probability $p$.  An exact relation on any bond-diluted lattice is
\begin{equation}
 f(p) = {{F}\over {gN}} = 1 - {z\over 2 g}[p-r(p)],
 \label{floppy}
\end{equation}
where  $F$ is the total number of floppy modes.
There are $N$ sites in the network, and $r(p)$ 
is the probability that a bond is redundant, i.e. its removal  does not
change the number of floppy modes in the system.  Eq.~({\ref{floppy}) 
demonstrates
that we can count either floppy modes or redundant bonds, and although
early work has emphasized counting floppy modes,\cite{T2,HT} they are 
equivalent. In Maxwell
counting, $r(p)$ is assumed to be zero for $p<p_m$ and the
rigidity threshold is taken to occur at $f(p_m)=0$, 
which leads to the {\it{Maxwell estimate}} of the rigidity threshold, 
\begin{equation}
p_m = 2g/z.
\end{equation}
Within this approximation it is evident that
 
\begin{eqnarray} 
f(p) &=& 0 \quad {\rm{for}} \quad p\ge p_m \nonumber \\
f(p) &=& 1-p/p_m \quad {\rm{for}} \quad p \le p_m, 
\end{eqnarray}
so that $f(p)$ changes slope at $p_m$. We emphasize that Maxwell 
counting  is 
not correct, but does provide a useful initial  approximation.  In 
reality 
$f \ne 0$ 
for $p \ge p_m$ and is non-zero right up to $p=1$ as there is always a 
small 
probability of having a floppy inclusion if $p \ne 1$.  This is a {\it 
Lifshitz} type of argument.\cite{L}

We note that $f(0) = 1$ and $f(1) =0$, and hence from 
Eq.~(\ref{floppy}) we have $r(0) = 0$ and 

\begin{equation}
r(1) = 1- {{2g}\over{z}}.
\label{rr}
\end{equation}
Equivalently, the following relation 
must be satisfied

\begin{equation}
\int_{0}^1 f^{(1)} dp = -1.
\label{minus}
\end{equation}
These consistency conditions remove boundary constraints from the
calculation and hence enable prediction of {\it bulk} critical 
behavior.
This is a key new element 
of the Cayley tree theory, which was lacking in MDL, and which we now 
refer to as the bulk or Bethe lattice solution. 
In particular it allows us to locate the bulk critical point
$p_c$ when the transition is first order.
 In MDL, the point at which the
finite real solution to the mean field equations ceased to
exist was identified as the critical point.  Here we label
that point as $p_s$ and reinterpret it as the spinodal point in
analog with thermally driven first-order transitions.  
 
We now show that the number of floppy modes acts as a free
energy in both connectivity and rigidity problems.  To demonstrate 
this, 
we recall the relation developed by Jacobs and Thorpe\cite{JT2}

\begin{equation}
f^{(1)}(p) = -{z\over 2g} \left(1-{N_0\over N_B}\right) = -{z\over 2g} 
{N_I\over 
N_B}
\label{Don}
\end{equation}
where $f$ is the number of floppy modes per degree of freedom, 
$N_0$ is the number of overconstrained bonds on the
lattice, $N_I$ is the number of isostatic  bonds on the
lattice, and 

\begin{equation}
N_B = {{pzN} \over {2}} = N_0 + N_I
\end{equation} 
 is the total number of bonds present on the
lattice, and the first derivative $f^{(1)} = \partial f/\partial p$. 
The relation 
(\ref{Don}) is 
obtained by 
{\it{removing}} a single bond, chosen at random, and ascertaining the 
probability that this
bond is in an overconstrained region
(in which case the number of floppy modes is unchanged) or else the 
number of 
floppy modes is reduced by 1. That is (for one removed bond) $\Delta F 
= (1 - 
N_0/N_B)$, and if
we now remove $\Delta N_B$ bonds, we have

\begin{equation}
{{\partial F} \over {\partial N_B}} = -\left(1-\frac{N_0}{N_B}\right),
\end{equation}
and the result (\ref{Don}) follows.
By comparing Eqs.~(\ref{floppy}) and (\ref{Don}), we see that the 
number of 
overconstrained bonds is given by the rate of change of the number of 
redundant 
bonds.

\begin{equation}
{{\partial r(p)}\over{\partial p}} ={{N_0}\over{N_B}}.
\label{over}
\end{equation}

We can use a similar argument to derive the important result  

\begin{equation}
f^{(2)}(p) ={z \over 2g} r^{(2)}(p)  \ge 0
\label{Convex}
\end{equation}
where the second derivative $f^{(2)} = \partial^2 f/\partial^2 p$.  We 
demonstrate 
this
as follows.
If we remove one bond, chosen at random, then the change in the number 
of 
overconstrained bonds is given by

\begin{equation}
\Delta N_0 = - \frac{N_0}{N_B}(1+\lambda) 
\end{equation}
where $\lambda \ge 0$ because when an overconstrained bond is removed, 
the 
number of overconstrained bonds is reduced by at least one, and 
sometimes by 
more.
If we remove $\Delta N_B$ bonds, we therefore have 

\begin{equation}
{{\partial N_0} \over {\partial N_B}} = \frac{N_0}{N_B} (1+\lambda)
\label{A}
\end{equation}
Using this, we find
\begin{equation}
{\partial\over \partial N_B} {N_0 \over N_B} 
= {1 \over N_B} \left({{\partial N_0} \over {\partial N_B}} - {N_0\over 
N_B}\right)
 = {\lambda  \over N_B}{N_0\over N_B}
\end{equation}
and hence using (9) and (10),
\begin{equation}
f^{(2)}(p) = \frac{\lambda z }{2 p g  } \frac{N_0}{N_B}  \ge 0
\label{B}
\end{equation}
establishing that $-f(p)$ is a convex function of the fraction of bonds 
present 
$p$.
 Note that we must ensemble average the results 
(\ref{A}) and (\ref{B}) so that $\lambda$ is to be interpreted as an 
ensemble 
averaged 
quantity.

Therefore we can use $-f(p)$ as a free energy, and if there is any
ambiguity the system will always be in the lowest free energy (maximum 
floppy 
modes) state. 
 The quantity 
$-f(p)$ is convex as required of a free energy, and we will refer to $f(p)$ 
interchangeably as 
the fraction of floppy modes {\it or} as the free energy.  Note that 
the 
proof above is 
given for arbitrary $g$ and $z$, but restricted to $b=1$ and bond 
dilution.  
Because $f^{(2)}(p)$ is the second derivative of a free energy and is 
positive
definite, we can regard it as a specific heat and it is calculated 
explicitly in subsequent sections.\cite{KTK}

\subsection{Cayley Trees}
\label{Sec:IIb}

Following MDL, we consider trees which have co-ordination number $z$  
with $g$ degrees of freedom per site, which form a Cayley tree network 
attached 
to 
a rigid boundary which we call a busbar. This is shown in Fig.~1.  
The busbar is not necessary for connectivity when 
$g=1$, but is of  vital importance for rigidity when $g \ge 2$.  We 
define 
$T^{n}_{0}$ to be
the probability that a bond on a {\it branch} $n$ levels away from the 
busbar  
is
part of the infinite rigid cluster.  In general, if
the sites of the tree have $g$ degrees of freedom,
 rigidity is transmitted to the next level of the tree provided 
 at least $g$ of the bonds to the lower level are occupied {\it and}
 provided that the sites at the ends of these bonds are rigid.
  This gives the recurrence relation

\begin{equation}
T^{n+1}_{0}= \sum_{k=g}^{z-1} {z-1 \choose k} 
(p T^n_{0})^k(1-p T^n_{0})^{z-1-k},
\label{rec1}
\end{equation}
where $T^n_{0}$ is the probability a bond $n$ levels from
the busbar is rigid.
If we take the thermodynamic limit (very large $n$), 
Eq.~(\ref{rec1}) iterates to a steady-state solution,
which we call $T_{0}$ and is given by

\begin{equation}
T_{0}= \sum_{k=g}^{z-1} {z-1 \choose k} 
(p T_{0})^k(1-p T_{0})^{z-1-k}.
\label{rec2}
\end{equation}
From this equation, we can find the probability of having a single  
degree of freedom 
with respect to the (distant) boundary

\begin{equation}
T_{1}= {z-1 \choose g-1} 
(p T_{0})^{g-1}(1-p T_{0})^{z-g} {\rm ,} \ \ g\ne 1.
\label{rec3}
\end{equation} 
and more generally for $l$ degrees of freedom with respect to the 
boundary 

\begin{equation}
T_{\l}= {z-1 \choose g-l} 
(p T_{0})^{g-l}(1-p T_{0})^{z-g+l-1}{\rm ,} \ \ \ 1\le l \le g.
\label{rec4}
\end{equation} 
Summing over all possibilities, we have the useful sum rule

\begin{equation}
\sum_{l=0}^{g} T_{l} = 1.
\label{sum}
\end{equation}
Equation (\ref{rec2}) is the self-consistent equation for 
the rigidity order parameter on bond-diluted Cayley trees.  
MDL also considered a more general class of problem in which
another degree of freedom $b\ge 2$ [the number of
constraints (or bars) between each pair of sites] is allowed. They also 
considered the case of site dilution, which is trivially 
related to bond dilution on 
a Cayley tree for $b=1$ only.

For trees with $b=1$, the probability that a  bond is overconstrained 
is
 $p T_{0}^2$, i.e. a 
bond is overconstrained if it is present {\it and} both of the sites at
its ends are already rigidly connected to the busbar.
Such bonds are the {\it{overconstrained}} bonds 
of the infinite cluster, as no other bonds are overconstrained far from 
the busbar, which is the bulk solution that we are seeking. Thus from 
(\ref{Don}) we have,

\begin{equation}
f^{(1)}(p) = -{z\over 2g} \left(1-{pT_{0}^2\over p}\right) = -{z\over 2g} 
(1-T_{0}^2).
\end{equation}
Differentiating a second time gives

\begin{equation}
f^{(2)}(p) = {{z}\over {g}} T_{0}{{\partial {T_{0}}}\over {\partial p}}.
\label{dprime}
\end{equation}
Since $T_0 \ge 0$, we must have solutions where

\begin{equation}
{{\partial {T_{0}}}\over {\partial p}} \ge 0
\end{equation}  
 to be acceptable. 
 
In a similar way the probability that a bond is
connected to the 
infinite rigid cluster via the busbar at one end, but
has one degree of freedom 
at the other is $2pT_0T_1$. Such bonds are the {\it{isostatic}} bonds 
associated with the infinite cluster.\cite{JT,JT2,MD} Thus the total 
probability that a bond, that is 
present,  belongs to the infinite cluster $P_{\infty}$ is 

\begin{equation}
P_{\infty} ={T_0}^2 +2T_0T_1.
\label{infty}
\end{equation}
 The probability that a bond that is present and belongs to the 
overconstrained part of the backbone is given by

\begin{equation}
P_{ov} = {T_0}^2.
\label{back}
\end{equation}

Since the number of redundant bonds $r(p)$ is zero  
for $p \le p_c$ and after finding  $r^{(1)}(p) = T_0^2$ from (20) using 
(1),
we must have,

\begin{equation}
r(p) = \int_{p_c}^p T_{0}^2 dp.
\label{1}
\end{equation}
In particular, combining Eq.~(25), with the consistency condition (4), 
we have
 
\begin{equation}
r(1) = \int_{p_c}^1 T_{0}^2 dp = 1 - {2g \over z}
\label{rr1}
\end{equation}
This is the key new component of the theory
developed here.  As in MDL, we solve
the tree equations for the order parameter $T_0$ but now we use 
Eq.~(26) to identify the rigidity threshold $p_c$. 
 Once we have $p_c$, the number of floppy modes at any
$p$ is given by [from Eq.~(1) using (\ref{1})],

\begin{equation}
f(p) = 1 - {z \over 2g} \left[p - \int_{p_c}^p T_{0}^2 dp\right] .
\label{2}
\end{equation}

The condition, Eq.~(26) may be regarded as 
a {\it{Maxwell condition}}, in analogy with that commonly used to 
locate the 
first order transition in thermodynamic systems.\cite{M}  

\section{Random bond model}

The {\it random bond model} consists of $N$
sites thrown down randomly, with each site joined at random to
$z$ other sites (without regard to distance) to form a $z$-coordinated 
network. Small rings appear with probability $O(1/N)$. It is convenient 
to show this network in a plane as in Fig.~2, where $z=3$.  The 
dimension of this model is only defined through the number of degrees 
of freedom $g$ associated with each site. In the Bethe 
lattice and in the random bond model, there are no loops in the 
thermodynamic limit $N \rightarrow \infty$ and so in a diagrammatic 
expansion, all terms agree, and the models are equivalent.
 For finite Cayley trees, 
there are loops involving the busbar as can be seen in Fig.~1.  
Likewise, there are loops in the random bond model as can be seen in 
Fig.~2.  So the equivalence is only in the bulk thermodynamic limit. 
We find this is a useful alternative viewpoint on the Bethe lattice 
limit.

We have numerically examined connectivity percolation having 
$g=1$, $z=3$, and the $g=2$, $z=6$
rigidity percolation case using the random bond model. These simulation 
results 
are 
compared 
with the 
corresponding exact Bethe lattice calculations in the next two 
sections. The pebble game\cite{JT,JT2,JH} was used to find the
number of floppy modes and the derivatives $f^{(1)}(p)$ and 
$f^{(2)}(p)$ for the $g=2, z=6$ case. For connectivity 
percolation with $g=1$, a similar pebble game algorithm 
can be constructed where only one pebble is assigned to a site.\cite{C2}
 In fact, as long as a site represents a rigid body
having $g$ degrees of freedom and not a point, the pebble game can
be straight forwardly generalized by assigning $g$ pebbles to each 
site.\cite{C1} 

For the connectivity percolation case, however, we actually used the 
$g=2$ pebble game, (two pebbles per site), by invoking an interesting
mapping that makes $g=2$ rigidity percolation equivalent to 
connectivity percolation. This mapping is valid for any generic 
network and is not limited to the random bond model. The mapping
consists of adding a single {\it ghost} site to the $N$ site network. 
The ghost site has $g=2$ as do all other sites in the network. An
additional $N$ bonds are placed between each of the $N$ sites in the 
network and the ghost site. With this additional ghost site and 
its associated bonds, the network consists of edge sharing triangles.
That is, any three sites in the network that are connected, 
are mutually rigid because they form edge sharing triangles
with the ghost site common to all triangles. 

The ghost site and its associated bonds, allow connectivity to be 
a sufficient condition for a set of sites to be mutually rigid. Since
rigid clusters are those that are connected, a one to one mapping  
between clusters in a $g=2$ network to a $g=1$ network is established. 
However, 
the number of floppy modes of the $g=2$ network (with the ghost site) will
be two more than the number of rigid clusters. These floppy modes
can be viewed as two translational motions for the ghost site, and
one rotational motion of each rigid cluster about the ghost site. 
By accounting for these two extra modes, an exact mapping is 
established. 
\begin{center}
\epsfxsize=2.5in          
\leavevmode\epsfbox{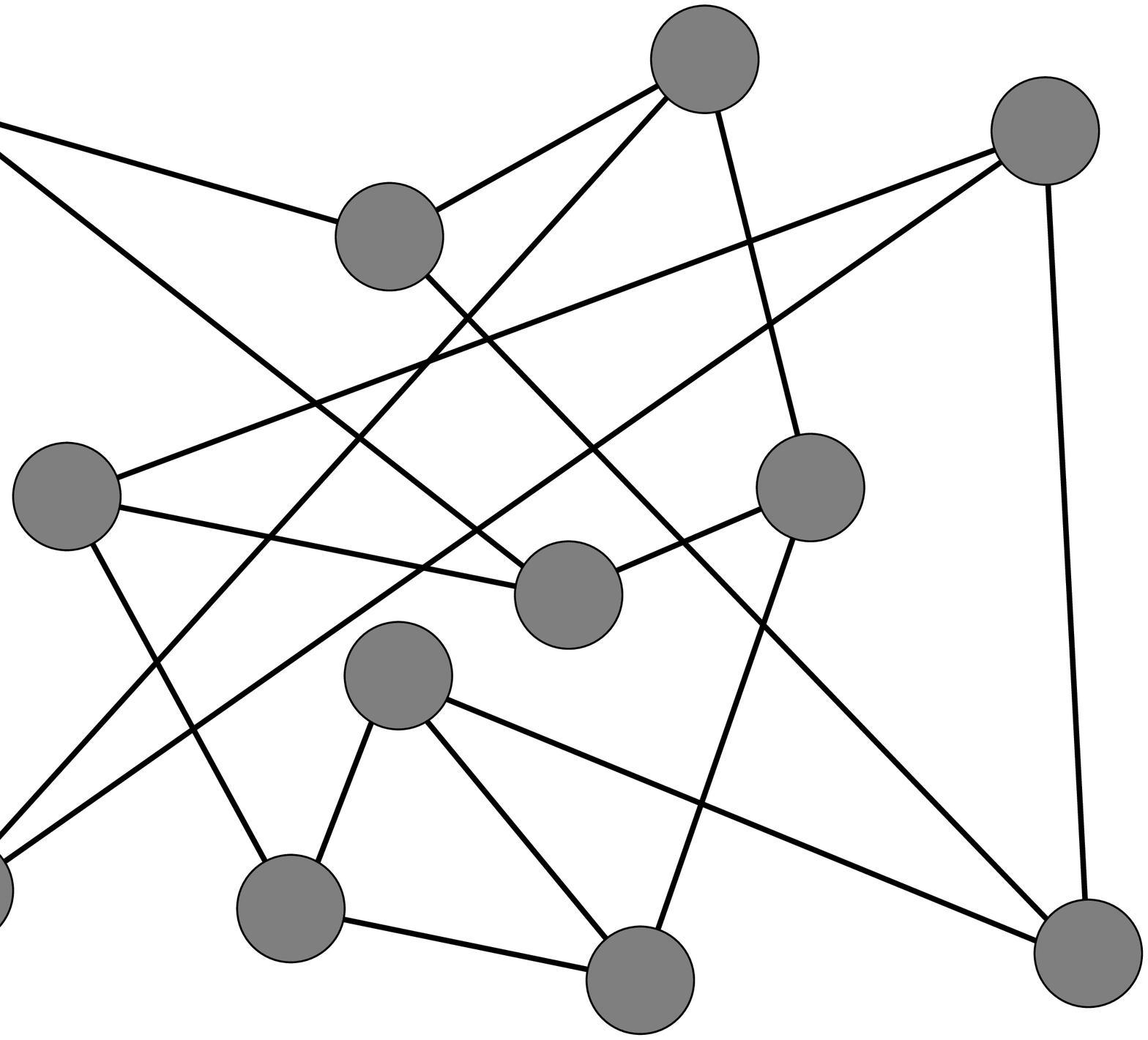}
\end{center}
\begin{figure}
\caption{A sketch of the random bond model with a single bar $b=1$ along 
the 
$z=3$ 
bonds 
from each site. The bonds are present 
with probability $p$.  Each site, shown as a shaded circle,  has $g$ 
degrees of 
freedom. Only $N=12$ sites are shown in this sketch.  In simulations 
using the pebble game, we use $N=262,144$.}
\label{graph55}
\end{figure}

\section{Connectivity Percolation}
The analysis in Sec.~II holds for any value of $g$.  For 
$g=1$ we 
recover and extend the familiar results of connectivity 
percolation\cite{FE,SA} 
for 
random 
bond 
dilution on a tree with coordination $z$.  In this section we use a 
general 
$z$ 
as much as possible, but then focus on $z=3$ 
to simplify the algebra when necessary. 
For $g=1$,  
Eqn.~(\ref{rec2}) reduces to 
\begin{equation}
T_0 = \left[1 - (1-pT_0)^{z-1}\right].
\label{T}
\end{equation}
Expanding near $p_c$ it is easy to show that the order parameter $T_0$ 
vanishes 
as

\begin{equation}
T_0 \sim \frac{2(p-p_c)}{p_c(1-p_c)} \quad {\rm{with}} \quad p_c = 
\frac{1}{z-1}.
\label{pc}
\end{equation}
The number of redundant bonds $r(p)$ can be found from (\ref{1}) by 
using 
(\ref{T}) 
to first calculate ${{\partial T_0} / {\partial p}}$ and then doing the 
integration to give

\begin{equation}
r(p) = T_0 \left[\left(\frac{z-2}{z}\right) (1+p-pT_0) - (1-p)\right]
\label{sc}
\end{equation}
where $T_0$ is obtained by solving the polynomial (\ref{T}). 
Notice that the {\it{fraction}} of bonds in overconstrained regions on 
the 
infinite cluster is 
given from Eqs.~(\ref{infty}) and (\ref{back}) and using $T_1=1-T_0$

\begin{equation}
\frac{P_{ov}}{P_{\infty}} = \frac{T_0^2}{{T_0}^2 + 2T_0T_1} = 
\frac{T_0}{2-T_0}.
\label{iso} 
\end{equation}
When $p=1$, we have $T_0 = 1$, so that the ratio (\ref{iso}) 
becomes unity, meaning that the entire tree is overconstrained (no 
dangling 
ends).
Near $p=p_c$, and using (\ref{pc}), the fraction of overconstrained 
bonds in 
the infinite cluster goes to zero linearly as
\begin{equation}
\frac{P_{ov}}{P_{\infty}} \sim \frac{p-p_c}{p_c(1-p_c)},
\label{iso2} 
\end{equation}
so that as the connectivity percolation transition is approached, the 
infinite
cluster  becomes completely isostatic.
For $z=3$ we can solve the quadratic equation (\ref{T}) to get

\begin{eqnarray}
T_{0} &=& 0 \quad {\rm{for}} \quad p \le p_c \nonumber \\
\label{TT}
T_{0} &=& \frac{(2p-1)}{p^2} \quad {\rm{for}} \quad p \ge p_c
\end{eqnarray}
where $p_c = 1/2$, and is shown in Fig.~3.
\begin{center}
\epsfxsize=2.5in          
\leavevmode\epsfbox{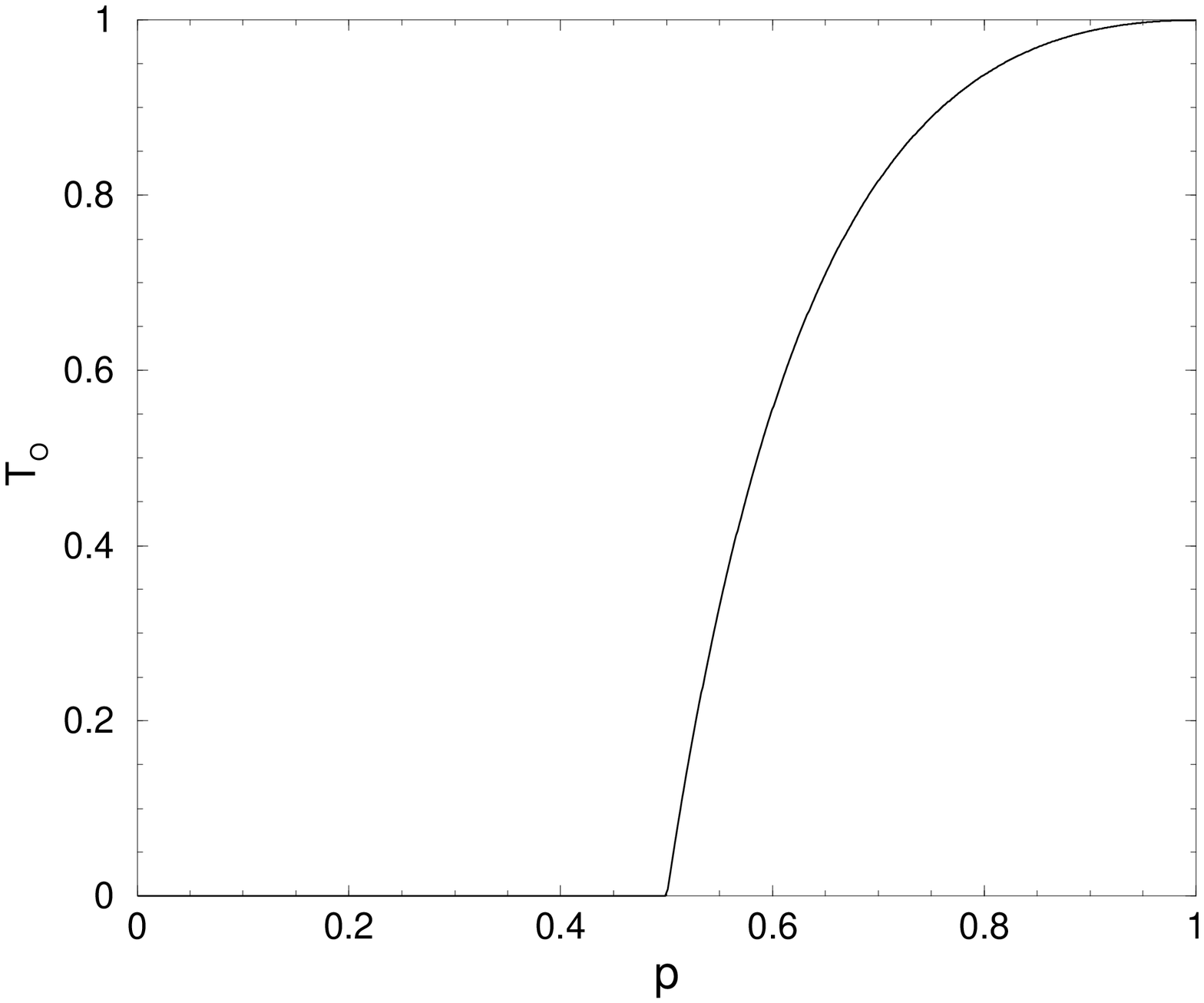}
\end{center}
\begin{figure}
\caption{The result for bond diluted connectivity percolation 
on a Bethe lattice for $T_0$ is shown for
co-ordination number $z=3$ and with $g=1$ degrees of
freedom per site. The probability that a bond
is present is $p$, and percolation occurs at $p_c = 0.5$.}
\label{graph2}
\end{figure}

 We can use (\ref{sc}) to check that we get 
the same value of $p_c$, by noting that we correctly reproduce 
r(1) = 1/3 [see Eq.~(\ref{rr})]. Therefore the transition is indeed
second order.  
Having found $p_c$, we find $f(p)$  and $r(p)$ 
from Eqs.~(\ref{1}) and (\ref{2}).

\begin{eqnarray}
f(p) &=& 1-\frac{3p}{2} \quad {\rm{for}} \quad p \le p_c  \nonumber \\
f(p) &=&  \left({{1-p}\over{p}}\right)^3 \frac{(3p-1)}{2} \quad 
{\rm{for}} \quad 
p \ge p_c
\label{ff}
\end{eqnarray}
and 

\begin{eqnarray}
r(p) &=& 0 \quad {\rm{for}} \quad p \le p_c   \nonumber \\
r(p) &=& \frac{1}{3} \left({{2p-1}\over{p}}\right)^3 \quad {\rm{for}} 
\quad p 
\ge p_c
\end{eqnarray}
and by differentiating (\ref{ff})

\begin{eqnarray}
f^{(1)}(p) &=& - \frac{3}{2} \quad {\rm{for}} \quad p \le p_c  
\nonumber \\
f^{(1)}(p) &=& -\frac{3}{2} \frac{(1-p)^2(p^2+2p -1)}{p^4} \quad 
{\rm{for}} \quad p 
\ge p_c
\end{eqnarray}
and differentiating again

\begin{eqnarray}
f^{(2)}(p) &=& 0 \quad {\rm{for}} \quad p \le p_c  \nonumber \\
f^{(2)}(p) &=& \frac{6(2p-1)(1-p)}{p^5} \quad {\rm{for}} \quad p \ge 
p_c.
\label{ffff}
\end{eqnarray}
\begin{center}
\epsfxsize=3.0in          
\leavevmode\epsfbox{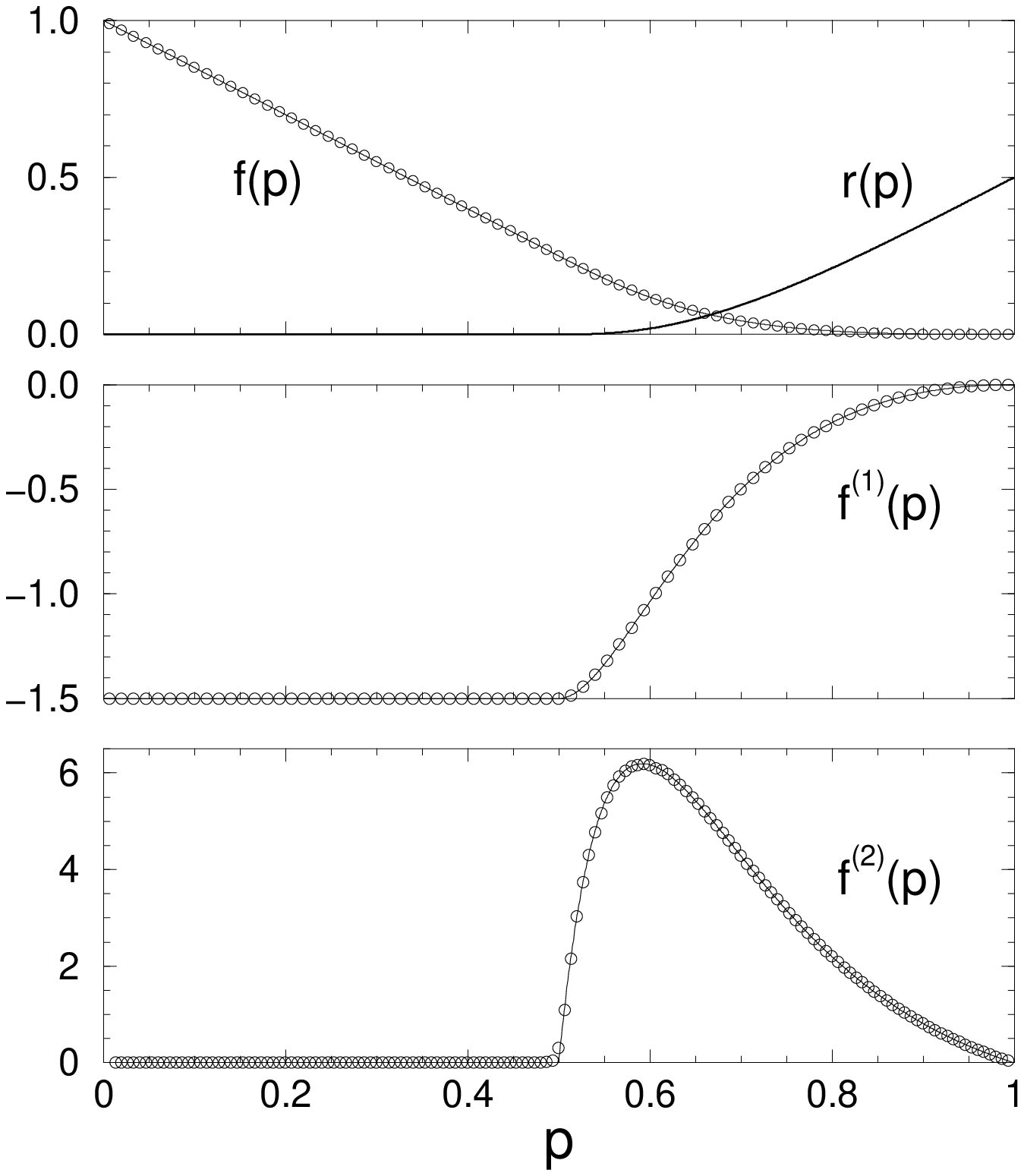}
\end{center}
\begin{figure}
\caption{Results for bond diluted connectivity percolation 
on a Bethe lattice for the number of floppy modes $f(p)$ and the first 
two derivatives. The number of redundant bonds $r(p)$ are also shown. 
The second derivative $f^{(2)}(p)$ acts as a specific heat for this 
problem.  Results are for
co-ordination number $z=3$ and with $g=1$ degrees of
freedom per site. The probability that a bond
is present is $p$, and percolation occurs at $p_c = 0.5$. The open 
circles are from computer simulations of the 
random bond model  using the pebble game as 
described in Sec.~III, using $N=262,144$ sites and 
averaging over 2,000 realizations. The Maxwell estimate [Eqs.~(2) and 
(3)]for 
the 
number of 
floppy modes is a straight line follows the $f(p)$ at small $p$.}
\label{graph3}
\end{figure}

These results (\ref{ff}) - (\ref{ffff}) are shown in Fig.~4.
We note that this second derivative, which is like a specific heat, is 
strongly 
peaked 
around $p= 1 - 1/\sqrt{6} = 0.59$, although it does go zero at $p_c 
=0.50$. 
Note also that 
$f, f^{(1)}$ and $f^{(2)}$ are all continuous at $p_c$, with $f^{(3)}$ 
being the first 
derivative to 
show a discontinuity.\cite{T2}

The number of  overconstrained bonds in the infinite cluster is $P_{ov}$ 
and is 
given 
from
(\ref{back}) by
\begin{equation}
P_{ov} = \left({\frac{2p-1}{p^2}}\right)^2 \quad {\rm{for}} \quad p \ge 
p_c
\end{equation}
and the fraction of bonds in the infinite cluster $P_{\infty}$ is given by 
(\ref{infty})
\begin{equation}
P_{\infty} = 1-\left(\frac{1-p}{p}\right)^4 \quad {\rm{for}} \quad p 
\ge p_c
\end{equation}
which has been derived by Essam and Fisher,\cite{FE}[see their Eq.~(35b)]
although notice they 
have an extra factor of $p$ as they normalized to all bonds, not just 
those 
{\it{present}}. These results are shown in Fig.~5.
The reader might be concerned that some overconstrained bonds 
are {\it{not}} associated with the infinite cluster.  This is not so on 
a Cayley 
tree in the asymptotic limit where only the infinite cluster 
reaches back to 
the busbar.  All other clusters are isolated and contain no 
overconstrained 
bonds, except finite rigid clusters attached to the busbar which are 
considered
to be surface effects and hence are ignored in the bulk or Bethe 
lattice solution.
\begin{center}
\epsfxsize=2.5in          
\leavevmode\epsfbox{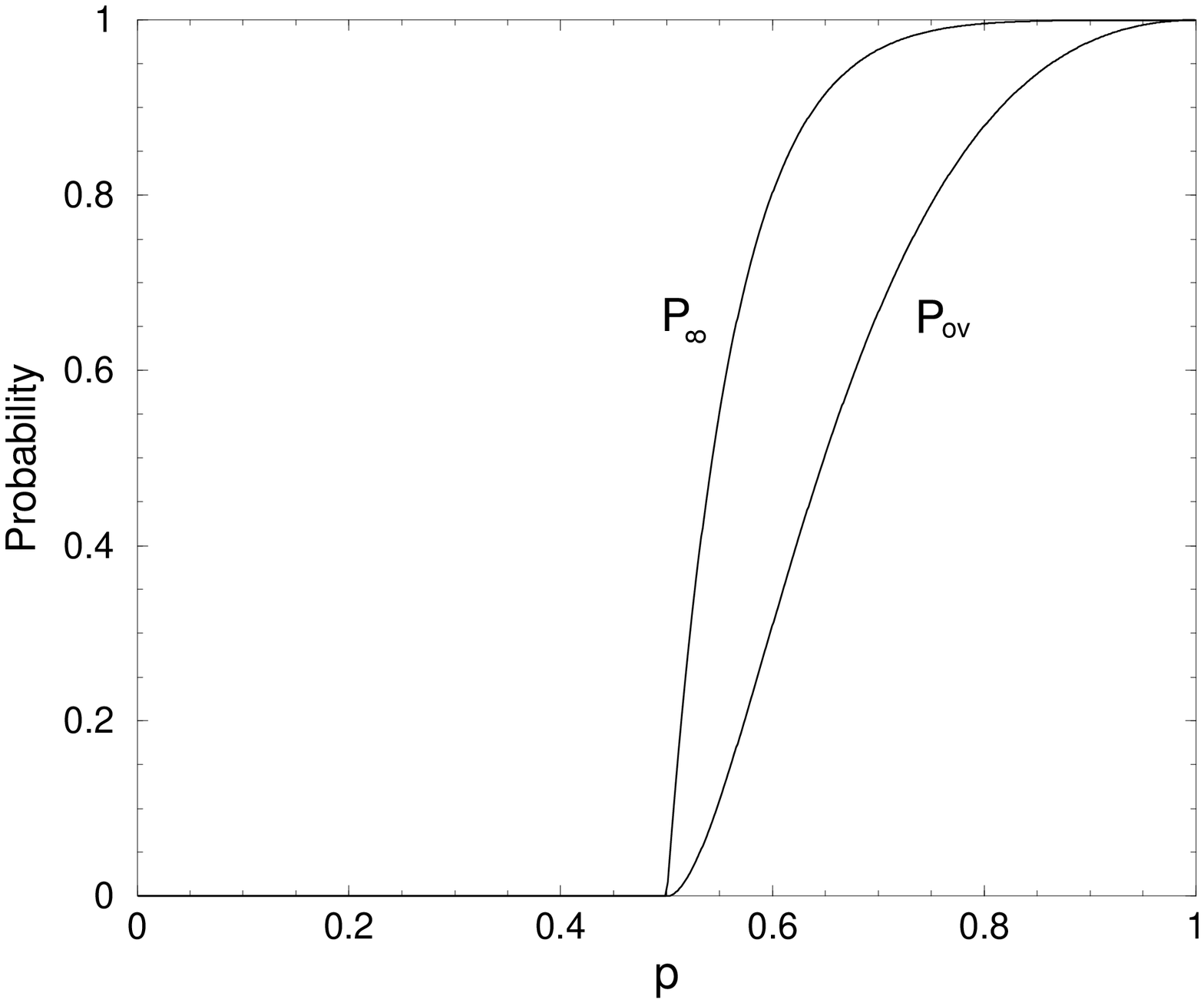}
\end{center}
\begin{figure}
\caption{Results for bond diluted connectivity percolation 
on a Bethe lattice for the probability of being on the
infinite cluster $P_{\infty}$ and the probability of being on the 
overconstrained part of the infinite cluster $P_{ov}$. Results are for
co-ordination number $z=3$ and with $g=1$ degrees of
freedom per site. The probability that a bond
is present is $p$, and percolation occurs at $p_c = 0.5$.}
\label{graph4}
\end{figure}

In the connectivity case $f^{(1)}(p)$ and the order parameters $T_0$, 
$P_{ov}$ and 
$P_{\infty}$  are continuous 
at the transition, and so connectivity percolation is of a conventional 
second 
order type, as indeed it is on real lattices like the triangular 
network.\cite{SA}

\section {Rigidity Percolation}

In the rigidity case we must numerically solve Eq.~(\ref{rec2}) 
for $T_0$ and hence find the other quantities of interest.
We first solve the self-consistent 
Eq.~(\ref{rec2}), by simple iteration, to find $T_0$, which is shown 
in Fig.~6 for $g=2$ and $z=6$. The analogue of the Maxwell 
construction (\ref{rr}) is then used to find $p_c$.  This is shown as 
the jump in Fig.~6 at $p_c = 0.656$.  The other possible solutions 
are ruled out by application of (\ref{rr}) and hence we have a
first order jump from a rigid to a floppy state.  The point at $p_s = 
0.603$ can be interpreted as a spinodal point.\cite{META}  The dashed line 
is 
unstable as can be seen from Eq.~(22).
Having found $T_0$ and $p_c$, we can find $f(p)$, $f^{(1)}(p)$, 
$f^{(2)}(p)$ and $r(p)$ as shown in Fig.~7. It can be seen that 
there are no redundant bonds for $p \le p_c$ and hence $f^{(1)}$ and 
$f^{(2)}$ 
are flat in this region, as also happens for connectivity percolation as 
shown 
in 
Sec.~IV.  In Fig.~7, we also show the results from the random bond model, 
discussed 
in 
 Sec.~III, and find that the results are equivalent as expected.  Note that 
with 
the 
random bond model, the first order transition occurs naturally and no 
Maxwell 
construction is needed, as the pebble game always finds the bulk {\it 
equilibrium} 
solution. 

\begin{figure}
 \begin{center}
\epsfxsize=2.5in          
\leavevmode\epsfbox{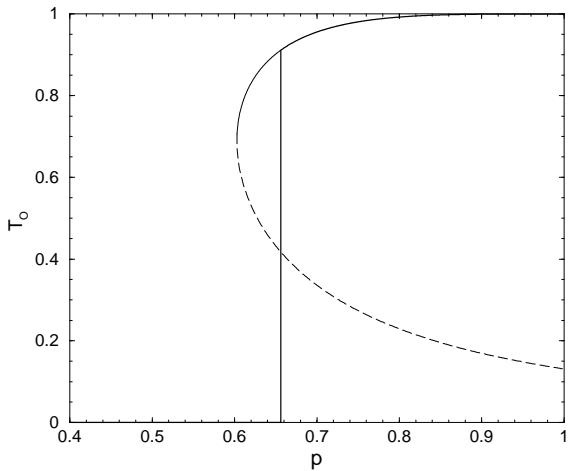}
\end{center}
\vskip -0.5cm
\caption{Results for bond diluted rigidity  percolation 
on a Bethe lattice for $T_0$ are shown for
co-ordination number $z=6$ with $g=2$ degrees of
freedom per site. The probability that a bond
is present is $p$ and percolation occurs at $p_c = 0.656$ shown by the 
vertical 
line.  The thin line 
extends out to the spinodal point at $p_s = 0.603$ and the dashed line 
shows the unstable solution.}
\label{graph5}
\end{figure}

In Fig.~8, we show results for 
$P_{\infty}$ and $P_{ov}$ from Eqs.~(23)and (24)
respectively.  It can be seen that most of the infinite cluster is 
overconstrained, 
even at 
the first order transition.

 In the connectivity case as shown in Figs.~3  and 5, the order
 parameter [ $T_0$, $P_{\infty}$ or $P_{ov}$] is continuous at $p_c$, 
whereas in 
the 
rigidity case as shown in Figs.~6 and 8 the order parameter 
 has a large first-order jump
 at $p_c$. The quantity $f^{(1)}(p)$, which acts like an energy, also has a 
large 
first order jump. One could argue that in the rigidity case,
 a metastable rigid state  exists 
 for any $p_c> p >p_s$.
 Note that for connectivity percolation, the boundary conditions are 
irrelevant, 
and 
no difference is found with or without a rigid 
busbar.  This is not so for rigidity percolation, where there are three 
real
 solutions\cite{CDL} for all $p_s<p<1$, with an {\it unstable}
 solution at finite $T_0$ existing in this regime as shown by the dashed 
line in Fig.~6. 
 In order for the stable finite $T_0$ solution to be found, the
 boundary rigidity must lie {\it above} the unstable fixed point.
 Any boundary rigidity below this value iterates to
 the $T_0=0$ solution.  This curious effect of the boundary condition in 
rigidity 
percolation, even on the bulk Bethe lattice solution, needs further study.
\begin{figure}
\begin{center}
\epsfxsize=3.0in          
\leavevmode\epsfbox{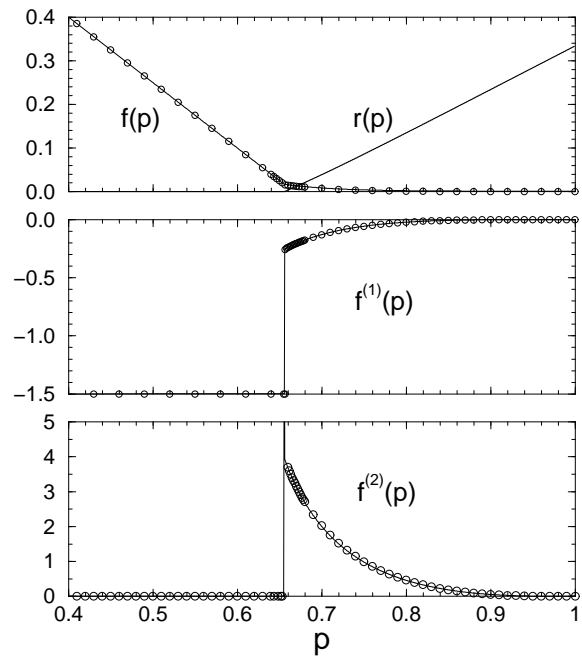}
\end{center}
\caption{Results for bond diluted rigidity percolation 
on a Bethe lattice for the number of floppy modes $f(p)$ and 
the first two derivatives. The number of redundant bonds $r(p)$ are 
also shown. The second derivative $f^{(2)}(p)$ acts as a specific heat 
for this problem.  Results are for
co-ordination number $z=6$ and with $g=2$ degrees of
freedom per site. The probability that a bond
is present is $p$ and percolation occurs at $p_c = 0.656$. The open 
circles are from computer simulations of the 
random bond model using the pebble game described in Sec.~III, using 
$N=262,144$ 
sites 
and 
averaging over 2,000 realizations. The Maxwell estimate [Eqs.~(2) and 
(3)]for 
the 
number of 
floppy modes is a straight line follows the $f(p)$ at small $p$.}
\label{graph6}
\end{figure}
\begin{center}
\epsfxsize=2.5in          
\leavevmode\epsfbox{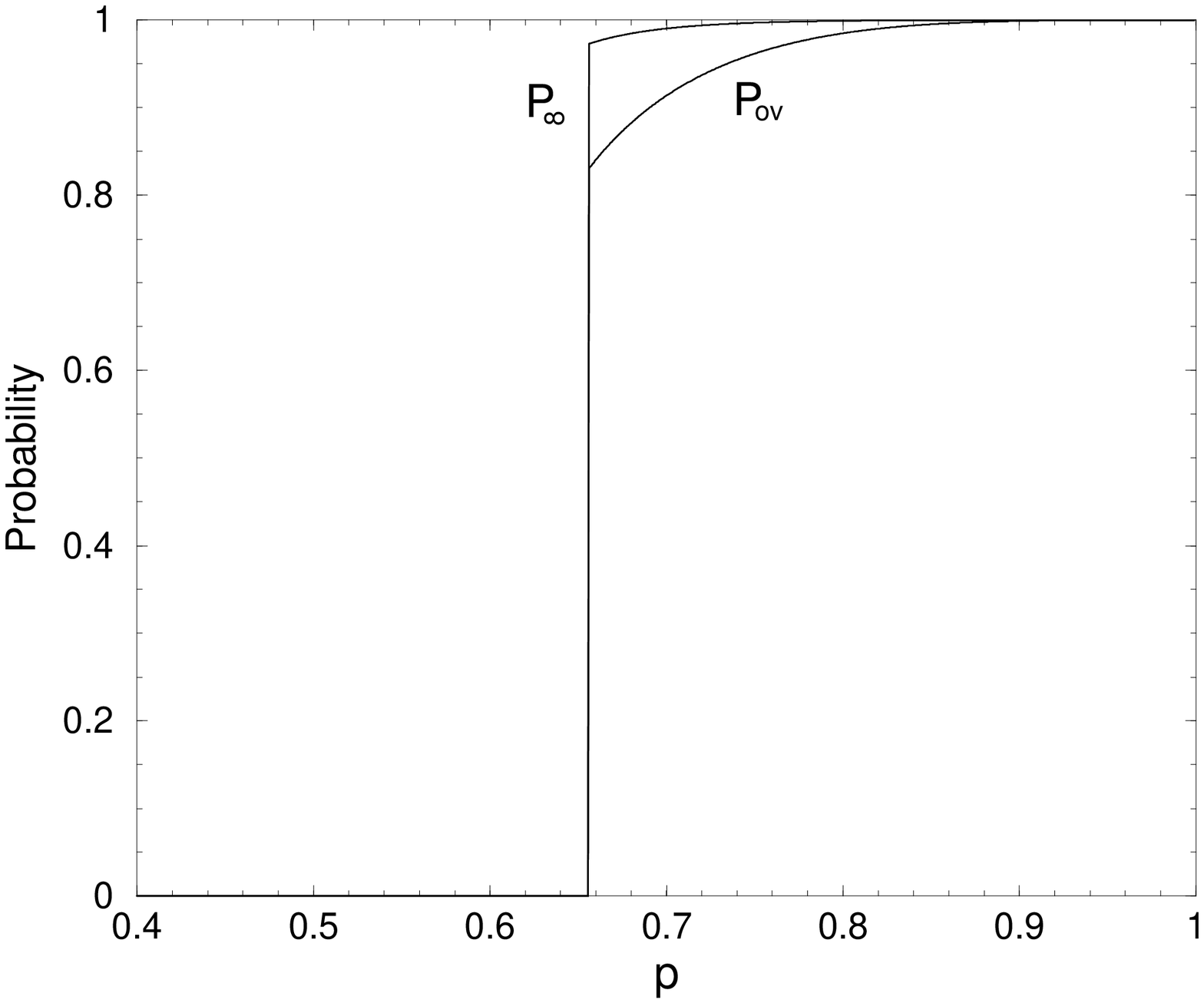}
\end{center}
\begin{figure}
\caption{Results for bond diluted rigidity percolation 
on a Bethe lattice for the probability of being on the
infinite cluster $P_{\infty}$ and the probability of being on the 
overconstrained part of the infinite cluster $P_{ov}$. Results are for
co-ordination number $z=6$  with $g=2$ degrees of
freedom per site. The probability that a bond
is present is $p$, and percolation occurs at $p_c = 0.656$.}
\label{graph7}
\end{figure}

  It is useful to compare the
results for $p_c$ with that predicted by Maxwell
 counting.  Maxwell counting would predict $p_m = 2g/z = 2/3$ in {\it
 both cases} studied here.  In the connectivity case this is markedly 
different
 than the exact result $p_c = 1/2$ because there are many 
 floppy modes at $p_c$ in that case as can be seen from Fig.~4.  In 
contrast, in 
the
 rigidity case, the Maxwell estimate is close to the
 Bethe lattice result $p_c = 0.655$  
 and there are rather few floppy modes at $p_c$ as shown in Fig.~7.  It is 
remarkable 
that Maxwell counting does so well in locating the position of {\it both} 
first 
order and second order\cite{STG,HT} rigidity transitions.
 
Results of calculations for a range of values of $g$ and
$z$ for bond diluted Bethe lattices are summarized in
table 1.  In all rigidity cases (i.e. $g>1$) 
there is a large jump in the order
parameter $T_{0}$, $P_{\infty}$ or $P_{ov}$ at the critical point $p_c$, 
although 
the 
jump does decrease with increasing  $z$ for a given $g$. 
On real lattices, there will also be a {\it Lifshitz tail} \ \cite{L} 
in the number of redundant bonds $r(p)$ [see Figs.~(4) and (7)] extending 
into 
the 
regime $p<p_c$ all the way to $p=0$.  Notice that the 
Bethe lattice result does correctly reproduce the Lifshitz tail in $f(p)$ 
in the 
region $p_c \le p \le 1$.

It is seen from Fig.~7
that the random-bond model is numerically equivalent to the
Bethe lattice results, as expected from our previous discussion. The 
pebble game is an exact algorithm for a particular network and gives the 
exact number of floppy modes (and the associated derivatives).  There 
is therefore no {\it hysteresis} in the results of the pebble game on 
the random bond model, and in these large samples, the system flips 
from the rigid state to the floppy state around $p_c=0.656$ as a 
single bond is removed in a given sample.   This is a 
very unusual situation as most numerical techniques, like Mont{\'{e}} 
Carlo encounter hysteresis near a first order transition.\cite{B}

\section{Large $z$ limit}

It is instructive to take the limit as the number of neighbors $z$ 
increases to 
infinity.  In order to make this limit sensible, it is necessary to change 
from 
the $p$ variable to $r=zp$, which is the mean coordination of a site.  For 
connectivity percolation, with $g=1$, Eq.~(16) becomes

\begin{equation}
T_0=1-e^{-rT_0}
\label{perc}
\end{equation}
which leads to a $r_c=zp_c=1$ which differs from the Maxwell estimate of
$r_m=zp_m=2$ by a factor $2$. Of more intertest to us is the case $g=2$ 
corresponding to rigidity, where Eq.~(16) becomes\

\begin{equation}
T_0=1-\left(1+rT_0\right)e^{-rT_0}.
\label{m1}
\end{equation}
Eq.~(\ref{m1}) can be solved for $T_0$ and the spinodal point found 
explicitly 
by differentiation. In this large $z$ limit, the analog of the {\it Maxwell 
condition} given in Eq.~(26) becomes

\begin{equation}
\int_{r_c}^\infty \left(1- T_{0}^2\right) dr = 2g - r_c.
\label{m2}
\end{equation}
Eqs.~(\ref{m1}) and (\ref{m2}) can be solved simultaneously to locate the 
first 
order transition at $r_c$, and the result is listed in Table 1.  We 
also show the results for the number of floppy modes $f$ at the transition 
and 
at the spinodal point, using Eq.~(27).  The values of $P_\infty$ and 
$P_{ov}$ 
are  obtained using Eqs.~(23) and (24) respectively.  It can be seen from 
Table 
1 that the values of all these quantities evolve smoothly as $z \rightarrow 
\infty$.

For larger values of $g$, we use Eq.~(16) to generalise Eq.~(\ref{m1}) to 
arbitary $g$

\begin{equation}
T_0=1-\left(\sum_{k=0}^{g-1}\frac{(rT_0)^k}{k!}\right)e^{-rT_0}.
\label{m3}
\end{equation}
Notice that Eq.~({\ref{m3}) reduces to Eqs.~({\ref{perc})
and ({\ref{m2}) respectively for $g=1$ and $g=2$.  Solving Eqs.~(\ref{m3}) 
and 
(\ref{m2}) simultaineously, we find the results for $g=3$ and $6$ listed in 
Table 1.  

By examining the equations in this section, we can find that as $g$ 
increases, 
the values of the mean coordination at the spinodal point $r_s$ and at the 
first 
order transition $r_c$  so that in the limit $g = \infty$, we have 

\begin{equation}
2r_s=r_c=r_m=2g
\end{equation}
and we recover the {\it Maxwell result} given in Eqs. (2) and (3).  Here 
$r_m$ is defined by $r_m = zp_m$.  It is interesting to note that the 
spinodal point is always distinct from the transition, and differs by a 
factor 2 in this limit. We obtained the limiting value of $r_s$ by 
carefully tracking Eq. (43) for large g, and find that the limit is 
approached very slowly. In this 
limit the metastable region between the spinodal and the first order 
transition 
exists between $r_s$ and $r_c$, and we have the Maxwell result which is 
also first order of course.

Note that in this limit $g \rightarrow {\infty} $, we have $T_0= 0$ for $r 
< r_c$ 
and $T_0= 1$ for $r > r_c$, so that the system is either {\it perfectly 
floppy} 
for $r < r_c$ when  there are no redundant bonds (this is true for arbitary 
$z$ 
and $g$) and now also {\it perfectly rigid} for $r > r_c$ in this limit 
with $z 
\rightarrow {\infty}$ and $g \rightarrow {\infty}$ when there are are no 
floppy 
regions.  This limit in which the system behaves homogeneously and the 
fluctuations are suppressed, may be regarded as the mean field limit for 
rigidity.

\section{Conclusions}

In this paper we have shown that the free energy is negative of the number 
of 
floppy 
modes.  This generalizes a previous result\cite{FK} which gives the free 
energy 
as 
the 
total number of clusters in connectivity percolation, where there is a 
single 
floppy 
mode associated with each isolated cluster.  In connectivity percolation, 
the 
free 
energy can also be found as the 
limit as $s \rightarrow 1$ of the $s$ state Potts model.\cite{P} No 
equivalent approach has been possible for rigidity.

We have used the free energy, and the associated Maxwell type construction, 
or 
self 
consistency condition, to show that the rigidity transition is always first 
order 
on 
Bethe lattices. It is necessary to use this construction in order to remove 
boundary constraints.  In 
previous work,\cite{MDL,CDL} a boundary dependent transition was found at 
the 
spinodal point $p_s$. As is expected from experience with thermodynamic 
transitions,
 metastability and boundary effects can be important\cite{META} in the
 regime $p_s<p<p_c$, and this needs further investigation.
  
 While rigidity concepts have been extensively applied
 to experiments in chalcogenide glasses\cite{HT,CT}, first
 order jumps in experimental quantities 
 have never been observed (although see recent Raman results in 
chalcogenide 
glasses\cite{XBB}).

\section{Acknowledgements}

We would like to thank A. R. Day for useful comments on the manuscript, the 
NSF 
for support under grant $\#$ DMR 9632182 and the DOE 
under 
contract  $\#$ DE-FG02-90ER45418.

\newpage

\begin{tabbing}
\hspace{0.5in}\= \hspace{0.5in}\= \hspace{0.9in}\= 
\hspace{0.9in}\= \hspace{0.9in}\=
 \hspace{0.6in}\= \hspace{0.6in}\= \hspace{0.6in}\= 
 \hspace{0.6in}\= \hspace{0.6in}\= \kill
 $g$  \> $z$ \> $r_m = zp_m$     \>  $r_s = zp_s$  \> $r_c = zp_c$ \> 
 $f(p_s)$  \> $f(p_c)$ \> $T_0(p_c^+)$  \> $ P_{\infty}(p_c^+)$   \> $ 
P_{ov}(p_c^+)$     \\

\> \\ 1  \> $z$ \>  2  \>  -\> ${z \over z-1}$ \> - \> ${z-2\over 2(z-1)}$ 
\> 0 
\>  0 \> 0 \\

\>\\ 2 \>   4 \>    4 \>  3.556 \>  4.000 \>  0.111 \>  0.000 \>  1.000 \>  
1.000 \>  1.000   
\>\\ 2 \>   6 \>    4 \>  3.617 \>  3.939 \>  0.096 \>  0.015 \>  0.912 \>  
0.973 \>  0.832   
\>\\ 2 \>   8 \>    4 \>  3.574 \>  3.867 \>  0.106 \>  0.033 \>  0.862 \>  
0.941 \>  0.743
\>\\ 2 \>   10 \>   4 \>  3.537 \>  3.816 \>  0.116 \>  0.046 \>  0.835 \>  
0.920 \>  0.698
\>\\ 2 \>   15 \>   4 \>  3.480 \>  3.743 \>  0.130 \>  0.064 \>  0.803 \>  
0.891 \>  0.645

\>\\ 2 \> $\infty$ \> 4 \>   3.352 \>  3.588 \> 0.162  \> 0.103 \>  0.749 
\>  
0.835 \>  0.561 \\
 
\>\\ 3 \> 6    \>   6 \>  5.009 \>  6.000 \>  0.165 \>  0.000 \>  1.000 \>  
1.000 \>  1.000

\>\\ 3 \> 8    \>   6 \>  5.167 \>  5.989 \>  0.139 \>  0.002 \>  0.984 \>  
0.996 \>  0.968
 
\>\\ 3 \> 10   \>   6 \>  5.206 \>  5.964 \>  0.132 \>  0.006 \>  0.964 \>  
0.987 \>  0.930
\>\\ 3 \> 15   \>   6 \>  5.217 \>  5.911 \>  0.131 \>  0.015 \>  0.936 \>  
0.969 \>  0.876
\>\\ 3  \> 25  \>   6 \>  5.200 \>  5.854 \>  0.133 \>  0.024 \>  0.913 \>  
0.950 \>  0.834

\>\\ 3 \> $\infty$ \> 6 \>  5.151 \>  5.755 \>  0.142 \>  0.041 \>  0.881 
\>  
0.919 \>  0.776\\

\>\\ 6 \> 12    \>  12 \> 9.600 \> 12.000 \>  0.200  \>  0.000 \>  1.000 \>  
1.000 \>  1.000

\>\\ 6 \> 15    \>  12 \> 9.360 \> 12.000 \>  0.220  \>  0.000 \>  1.000 \>  
1.000 \>  0.999
\>\\ 6 \> 25    \>  12 \>  9.646 \> 11.990 \>  0.196 \>  0.001 \>  0.994 \>  
0.997 \>  0.987
\>\\ 6 \> 50    \> 12  \>  9.785 \> 11.966 \>  0.185 \>  0.003 \>  0.985 \>  
0.990 \>  0.970
\>\\ 6 \> $\infty$ \> 12 \> 9.871 \> 11.930 \>  0.177 \>  0.006 \> 0.974 \> 
0.980 \> 0.949\\

\end{tabbing}

\hsize\textwidth\columnwidth\hsize\csname @twocolumnfalse\endcsname
{\bf Table 1 }  
 This table gives a summary of results for bond-diluted Bethe lattices with 
$g$ 
degrees of freedom per site which are joined by $b=1$ bars to $z$ 
neighbors. The 
Maxwell estimate for the transition $r_m=zp_m$ is given as well as the 
spinodal 
point 
$r_s=zp_s$ 
and the 
  critical point $r_c=zp_c$, where we mulitply these quantities by the 
coordination number $z$ to make them coordination numbers, that are 
semi-invariant. The number of floppy modes at the spinodal point is 
$f(p_s)$ 
and at the critical point is $f(p_c)$.  The jumps in the order parameters 
are   
given 
for $T_0$, and the  probability $P_{\infty}$
that a bond is present and is part of the infinite cluster, 
 and the probability $P_{ov}$ that a bond is present and is 
overconstrained.


\begin{thebibliography}{99}
\bibitem{KTK} An alternative form for the {\it specific heat} 
$(qd/dq)^2f(p) = 
q[f^{(2)}(p) - f^{(1)}(p)]$, where $q=1-p$ has been used by S. Kirkpatrick, 
Phys. 
Rev. Lett., {\bf 36}, 69 (1976) and M.F. Thorpe and S. Kirkpatrick, J. 
Phys. 
{\it 
A 
12}, 1835 (1979).  This form is also positive as can be seen from Eqs.~(6) 
and 
(10).
\end{thebibliography}
\end{document}